\documentclass{llncs}
\usepackage[utf8]{inputenc}
\usepackage[english]{babel}

\usepackage[numbers]{natbib} 
\usepackage[hidelinks]{hyperref} 

\usepackage{graphicx}
\usepackage{wrapfig} 
\usepackage{subfig}

\title{User Guidance for Interactive Camera Calibration}

\author{Pavel Rojtberg \inst{1, 2}}

\institute{Fraunhofer IGD, Darmstadt, Germany \\ \email{pavel.rojtberg@igd.fraunhofer.de} \and TU Darmstadt, Germany}

\date{}

\begin{document}

\maketitle

\begin{abstract}
For building a Augmented Reality (AR) pipeline, the most crucial step is the camera calibration as overall quality heavily depends on it.
In turn camera calibration itself is influenced most by the choice of camera-to-pattern poses – yet currently there is only little research on guiding the user to a specific pose.
We build upon our novel camera calibration framework that is capable to generate calibration poses in real-time and present a user study evaluating different visualization methods to guide the user to a target pose.
Using the presented  method even novel users are capable to perform a precise camera calibration in about 2 minutes.
\end{abstract}


\section{Motivation}


Camera calibration in the context of Augmented Reality (AR) is the process of determining the internal camera geometrical and optical characteristics (intrinsic parameters) and optionally the position and orientation of the camera frame in the world coordinate system (extrinsic parameters).
The performance of the 3D vision algorithms which AR builds upon directly depends on the quality of this calibration.
Furthermore, calibration is a recurring task that has to be performed each time the camera setup is changed. Even if a camera is replaced by an equivalent from the same series, the intrinsics will vary due to build inaccuracies.

\begin{figure}
\includegraphics[width=0.49\textwidth]{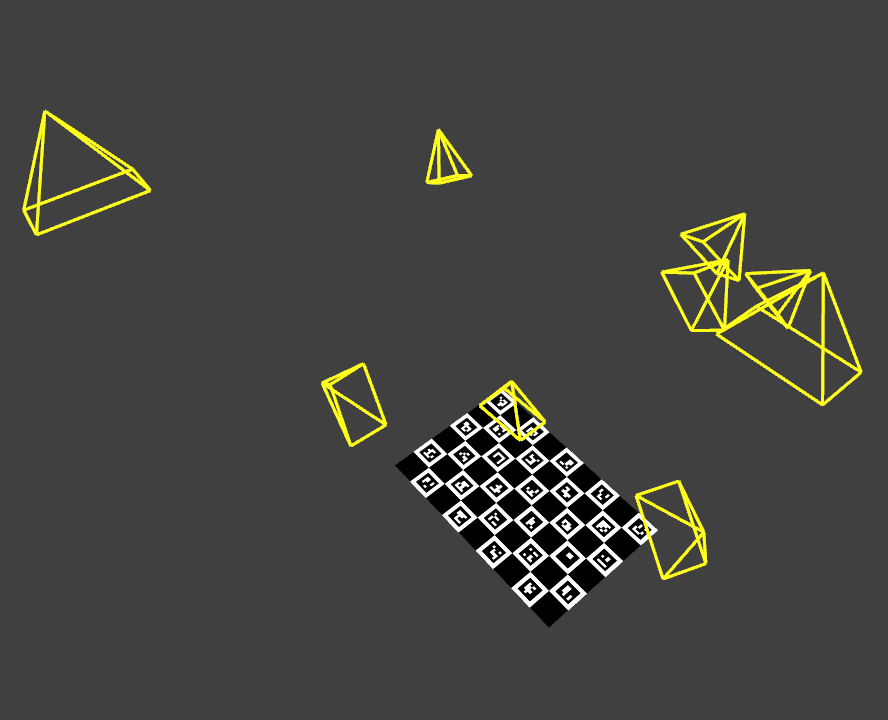}
\includegraphics[width=0.49\textwidth]{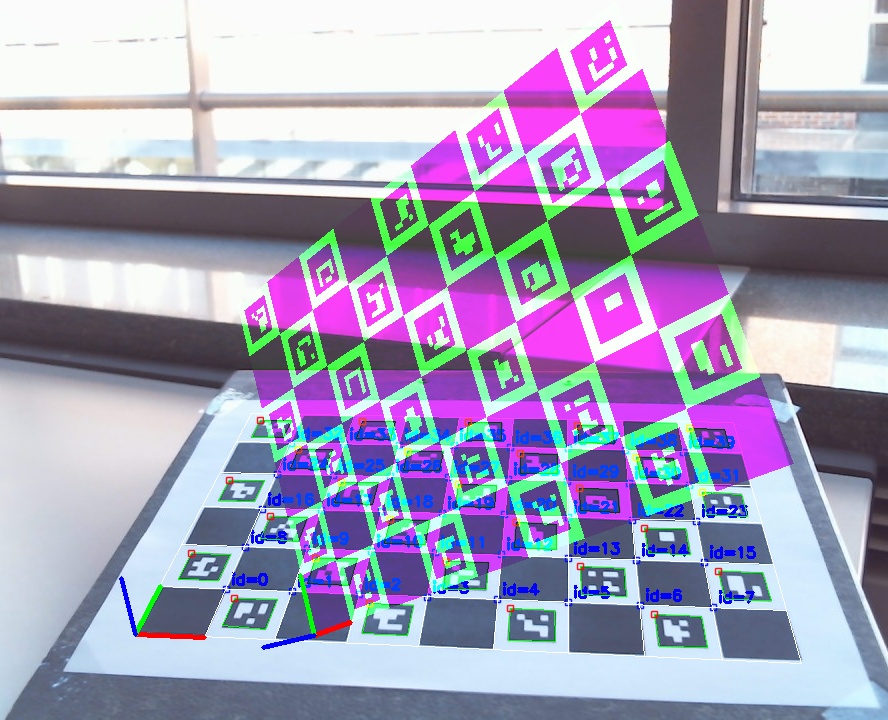}
\caption{Exemplary set of 9 target poses and the user guidance overlay, projecting for the bottom right camera.}
\label{fig:init_overlay}
\end{figure}

The prevalent approach to camera calibration is based on acquiring multiple images of a planar pattern of known size \cite{zhang2000flexible}.
In contrast to 3D calibration objects that were used earlier, 2D patterns are easy to obtain as conventional printers can produce them at high precision.

The pattern is then used to establish correspondences between known 3D world points and measured 2D image points.
The point correspondences form a over-determined system of equations which constrains the camera model.

However, due to the projective nature of the transform multiple images must be acquired from different poses. Here, special pose configurations \cite{sturm1999plane} that lead to an unreliable solutions and should be explicitly avoided.

Therefore a user interface is desirable which guides users through the calibration process. The guidance allows to select to a minimal set of "good" frames that result in a fast and reproducible calibration.

\section{Background}

\begin{figure}
\subfloat[Chessboard] {
\includegraphics[width=0.49\textwidth]{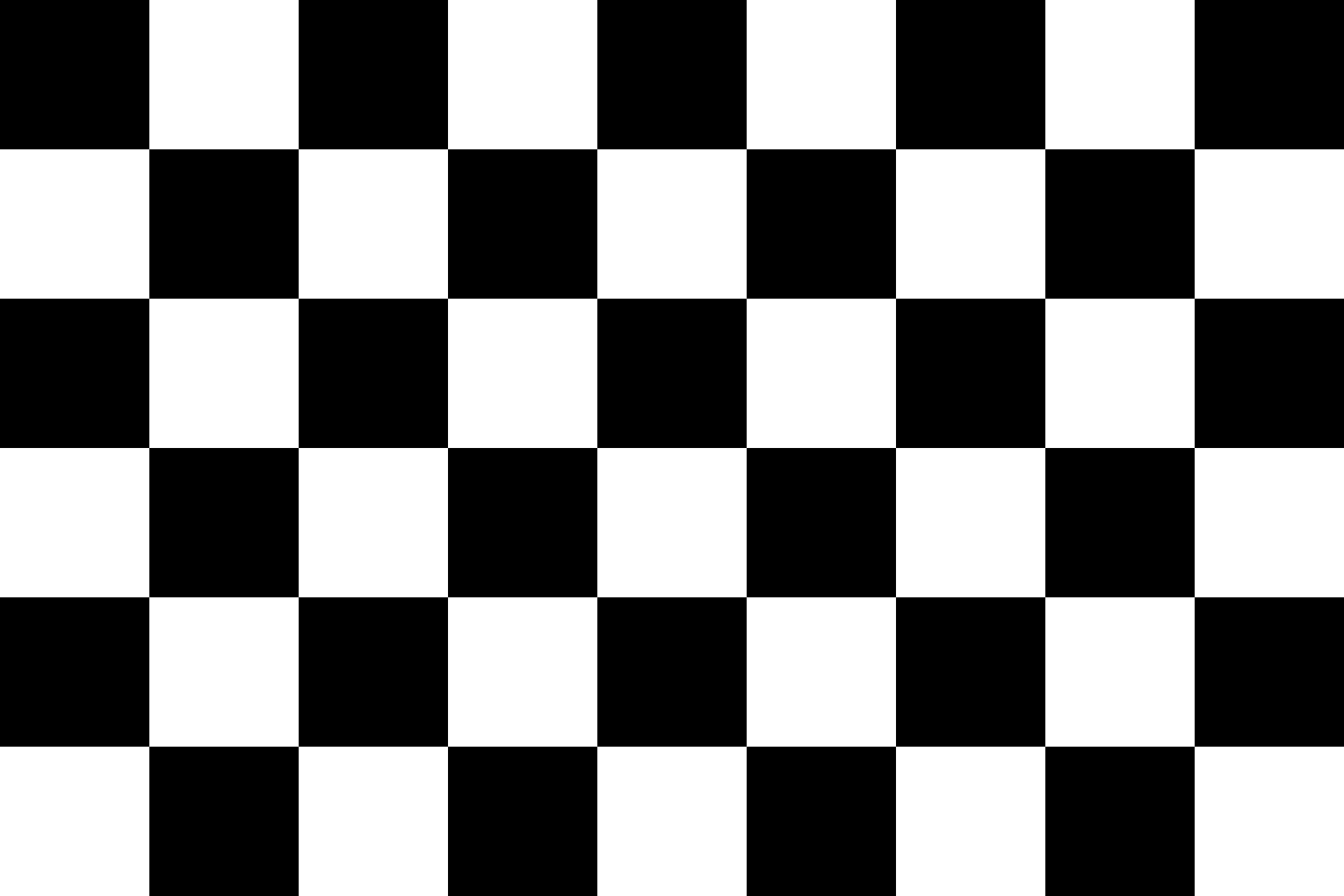}
\label{fig:chess-board}
}
\subfloat[ChArUco pattern] {
\includegraphics[width=0.49\textwidth]{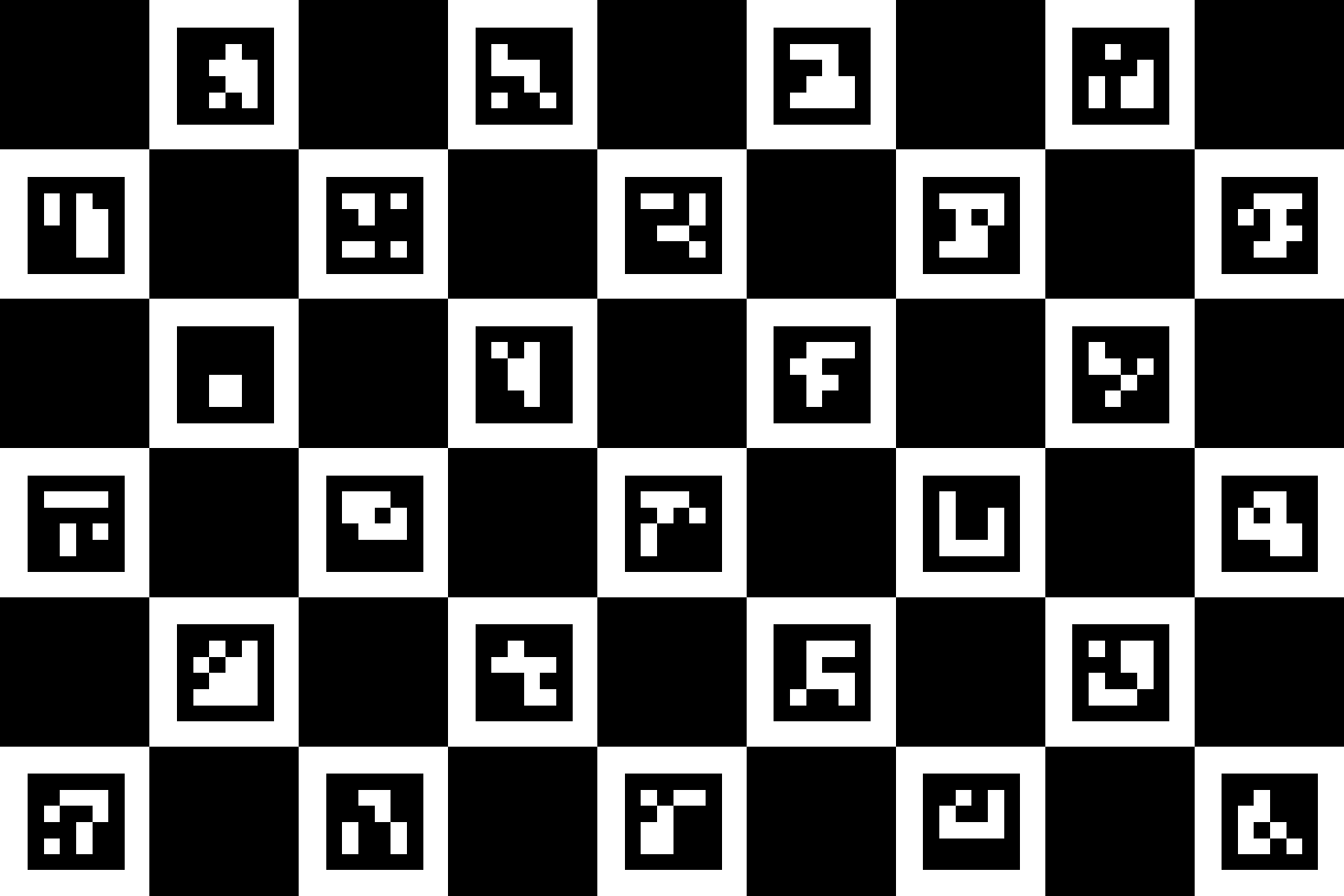}
\label{fig:charuco-board}
}
\caption{Common planar calibration patterns}
\label{fig:boards}
\end{figure}

The first step to camera calibration is to detect the calibration pattern. Typically chessboard patterns (Fig. \ref{fig:chess-board}) are used, which have strong gradients that can be detected even under difficult lighting conditions. Additionally, the 2D points can be located with sub-pixel accuracy.

However, the detection process usually involves the time consuming task of ordering the detected rectangles to a canonical topology. This slows down frame acquisition to below 30Hz and impedes the interactivity of the method.
Furthermore the board needs to be fully visible for the corner identification to work. 

Therefore new methods interleave fiducial markers \cite{garrido2014automatic} within the chessboard pattern (Fig. \ref{fig:charuco-board}). The markers encode an unique id and are designed to be rotationally invariant - hence the detection of a single marker allows deducing the location and orientation of the whole board. However the marker positions become imprecise at steep view angles. Hence only chessboard corners are used, which generate measurements at sub-pixel accuracy.

\begin{figure}
\subfloat[The ROS toolbox showing the variance in position and size.] {
\includegraphics[width=0.49\textwidth]{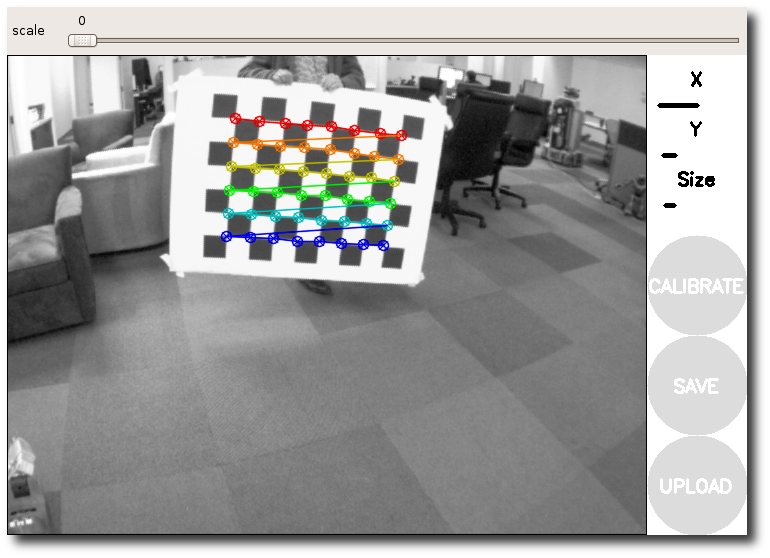}
}
\subfloat[OpenCV showing the current screen coverage] {
\includegraphics[width=0.49\textwidth]{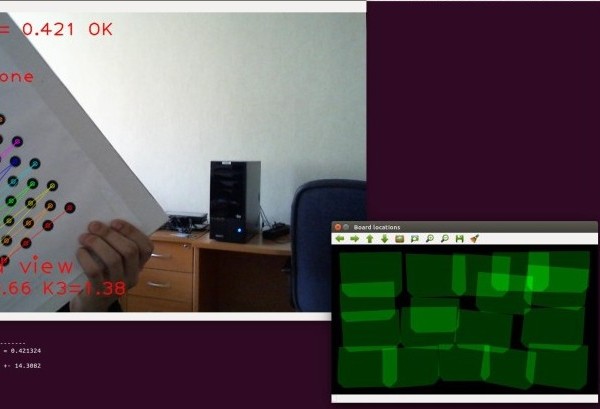}
}
\caption{User interfaces of popular, non-interactive, systems.}
\label{fig:others}
\end{figure}

The second step in calibration is to capture a calibration set of multiple images. This set needs sufficiently constrain the camera model for the calibration to succeed. For instance the pattern vies must not be all parallel to the image plane. As both pattern distance and camera focal length ("zoom level") are estimated simultaneously, there is no unique solution in this case.
Consequently popular calibration toolboxes like ROS\footnote{\url{http://wiki.ros.org/camera_calibration/Tutorials/MonocularCalibration}} or OpenCV \footnote{\url{https://docs.opencv.org/master/d7/d21/tutorial_interactive_calibration.html}} impose some heuristics on pose variance or screen space coverage to alleviate the problem (see figure \ref{fig:others}).

As these systems are not capable of generating pose suggestions, their user interfaces only visualize statistics about the data captured so far. The user is responsible to reason about an optimal next pose that would improve on the imposed heuristics.
Furthermore the unreliable pose configurations are not explicitly addressed --- therefore degraded performance is still possible.

\begin{figure}
\subfloat[Target pose wireframe and real-board overpaint as used by \cite{richardson2013iros}] {
\includegraphics[width=0.49\textwidth]{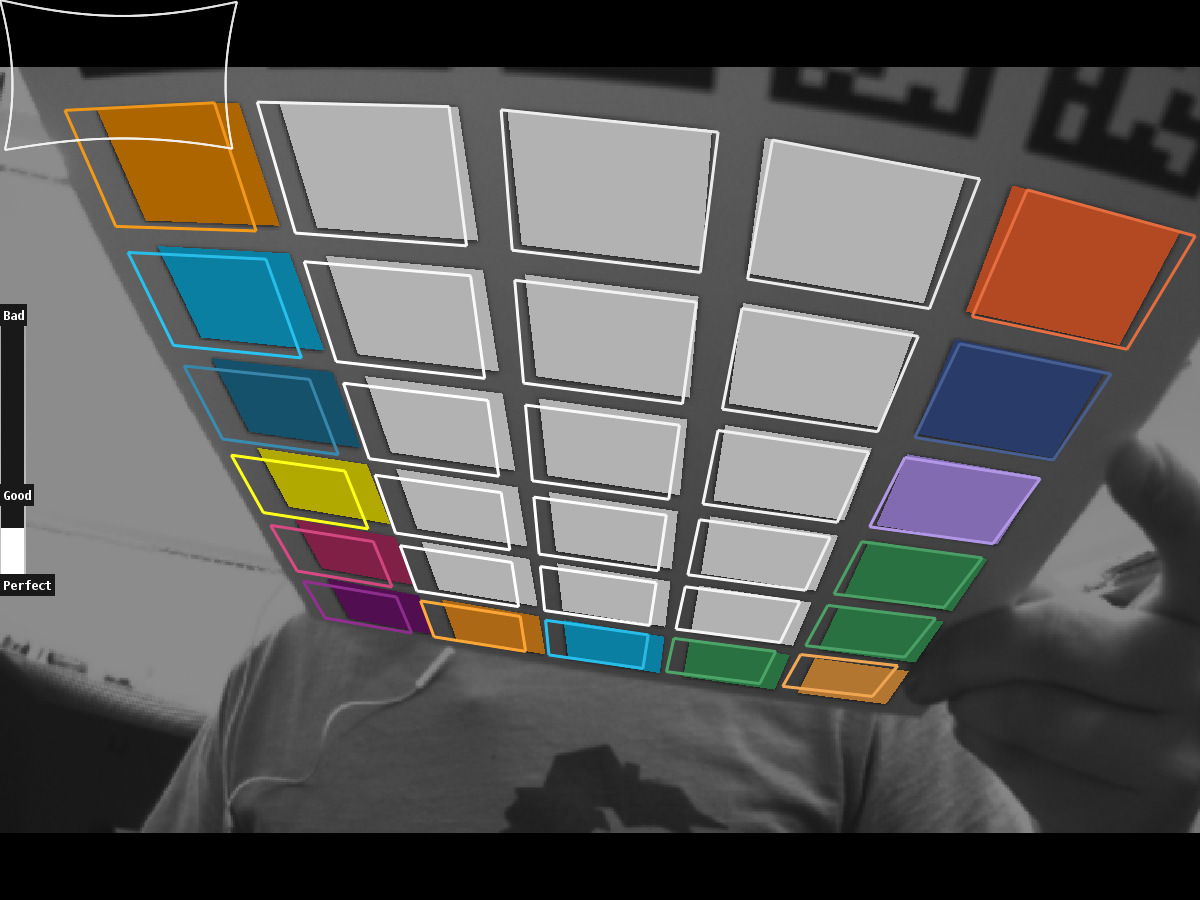}
}
\subfloat[Target pose projection as used by \cite{rojtberg2018}] {
\includegraphics[width=0.49\textwidth]{figure/overlay2.jpg}
}
\caption{User guidance overlays used by interactive calibration systems}
\label{fig:overlays}
\end{figure}

In contrast, new calibration systems \cite{richardson2013iros, rojtberg2018} are capable to guide users to specific target poses by displaying an overlay (see figure \ref{fig:init_overlay}).
This explicitly avoids unreliable configurations and reduces intrisic cognitive load \cite{chandler1991cognitive}.
While both \cite{richardson2013iros, rojtberg2018} performed user surveys, they merely showed operability of their methods by novice users.

Additionally the user interfaces implemented by each method are very different. \cite{rojtberg2018} only display highlighted projection of the real pattern to tag the target pose, while \cite{richardson2013iros} display an abstractly colored, wireframe of the board at the target pose and additionally overpaint the real board with squares of matching color (see Figure \ref{fig:overlays}).

Therefore this work focuses on the question how which user interface is best suited to guide users to specific calibration poses.
At this we take the specific geometric properties of the calibration problem into account, namely:
\begin{itemize}
\item Only the relative pose between camera and pattern matters
\item The pattern can be arbitrarily flipped horizontally and vertically.
\end{itemize}

Indeed, these properties make the calibration guidance significantly different from typical AR guidance use-cases where a pose needs to be matched exactly.

\subsection{Calibration Poses}

In general a rigid pose has six degrees of freedom (DOF); yaw, pitch, roll for the orientation and the three dimensional position.
However the underlying algorithm \cite{rojtberg2018} generates more restricted poses, based on the calibration objective. These fall in the following two categories: 

\begin{figure}
\subfloat[Intrinsic calibration pose] {
\includegraphics[width=0.49\textwidth]{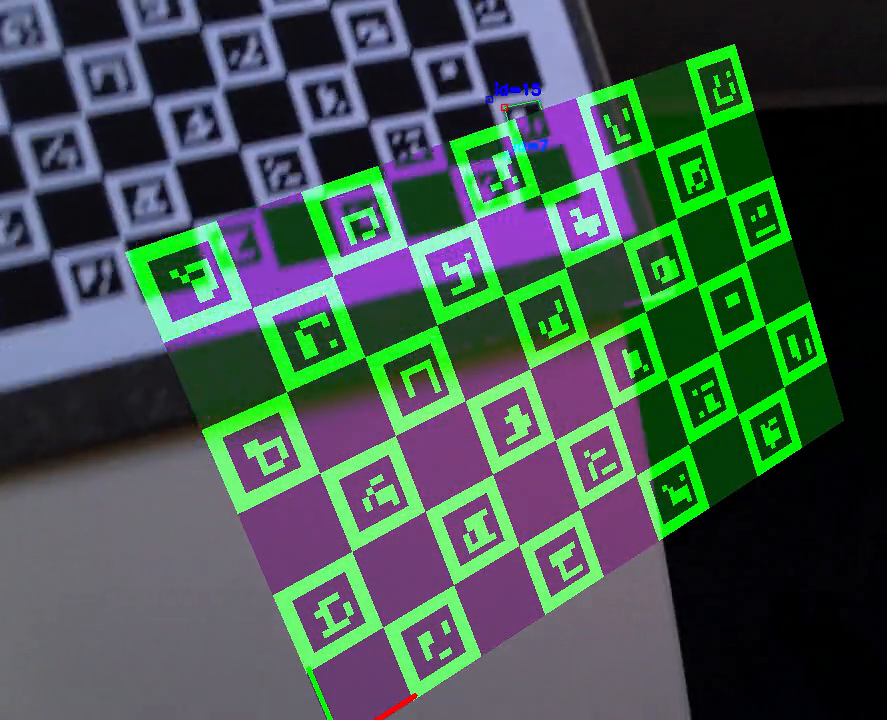}
\label{fig:intrpose}
}
\subfloat[Distortion calibration pose] {
\includegraphics[width=0.49\textwidth]{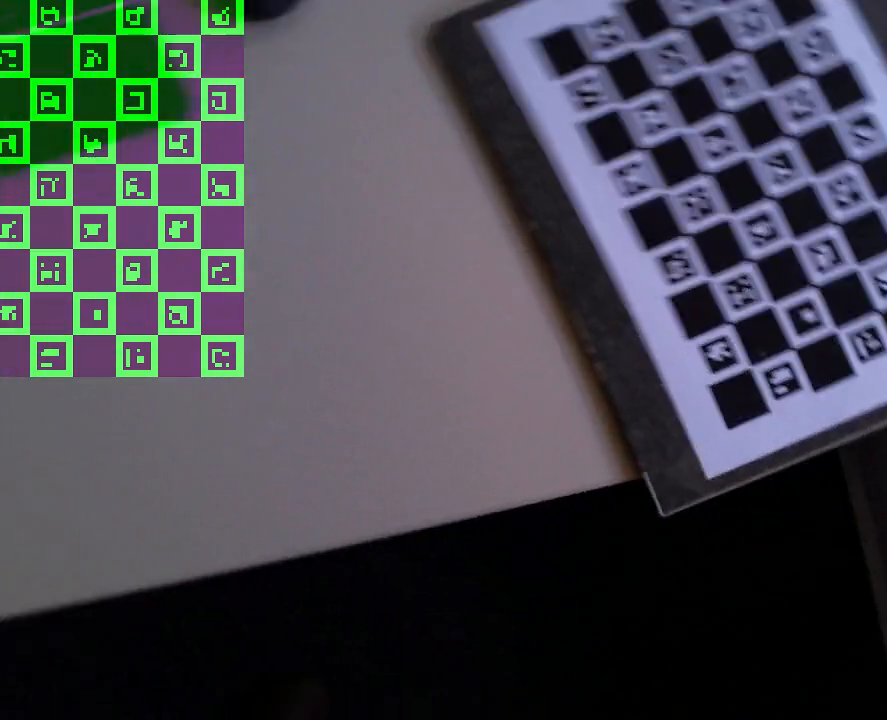}
\label{fig:distpose}
}
\caption{Exemplary view from the two pose categories}
\label{fig:poses}
\end{figure}

\paragraph{Intrinsic calibration pose} To estimate the intrinsic camera parameters, the goal is to maximize the angular spread of the measurement points. Here the pattern is placed in the central image region and tilted along one primary axis. Additionally the board needs to be tilted and rotated along the remaining axes to avoid ambiguous configurations (see Figure \ref{fig:intrpose}). Therefore there are only three rotational DOF.

\paragraph{Distortion calibration pose} To estimate the lens distortion parameters, the pattern must be placed in regions with highest distortion which are typically the corners. Here a parallel view is used and the distance and relative position changes (see Figure \ref{fig:distpose}). Therefore there are only three positional DOF.

Therefore a user only ever has to change 3 DOF when starting from a central, parallel view on the pattern.

\section{Method}

To evaluate different user guidance options, we performed two user surveys, measuring the time the users required to match a series of target poses. The participants were students and co-workers at our lab. Most of them had never performed a camera calibration before and all users were using the tool for the first time. The pose sequence was given by our system \cite{rojtberg2018}.

The only instruction given was that the calibration pattern should be matched with the displayed overlay.

We triggered the time measurement only after the first target pose was reached. This explicitly discards the time the users needed to accommodate to the calibration scenario and the system setup.

For each question a separate survey was performed. The surveys were several months apart time-wise. Hence there is no overlap of participants and the pose setup varies slightly.

\subsection{Relative motion survey}

The goal of the first survey was to determine whether moving the camera or moving the calibration pattern is preferable. This takes advantage of the fact that only the relative orientation and translation between camera and pattern matters. Therefore we evaluated the following two scenarios:

\begin{enumerate}
\item Fixing the camera position at the screen and let the user move the pattern like in front of a virtual mirror.
\item Fixing the pattern position and let the user move the camera in a first-person-view like fashion.
\end{enumerate}

There were 5 participants in this survey which successively tried both options. To exclude the effect of familiarization we randomized the the order of the options. The user guidance consisted only of the target pose overlay as shown in Figure \ref{fig:poses}. There were 9 target poses that had to be matched.

\subsection{Pattern appearance survey}

\begin{figure}
\subfloat[The default "chessboard" pattern] {
\includegraphics[width=0.49\textwidth]{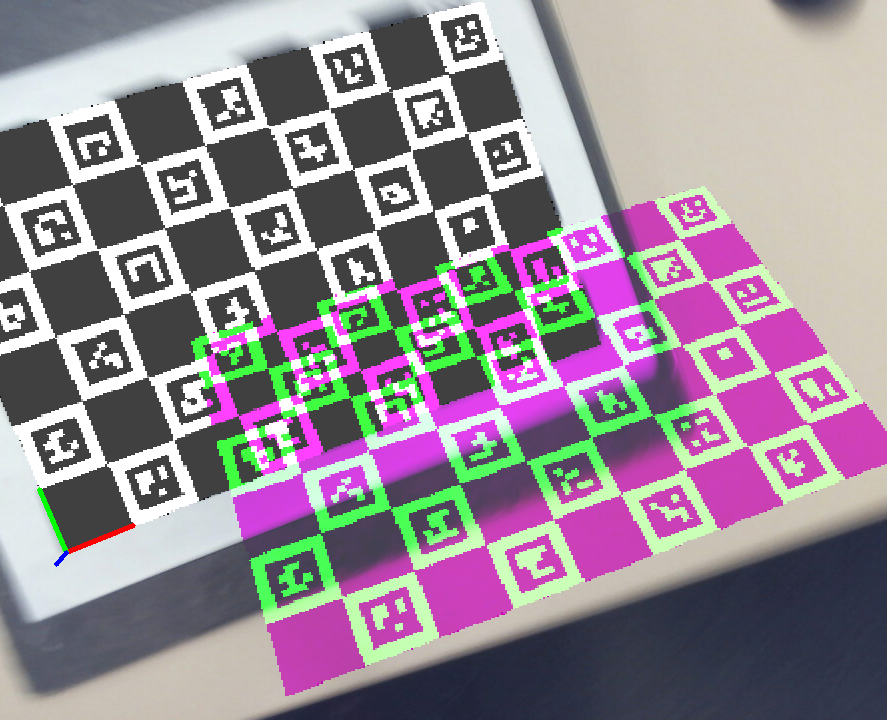}
}
\subfloat[The "quadrille paper" pattern] {
\includegraphics[width=0.49\textwidth]{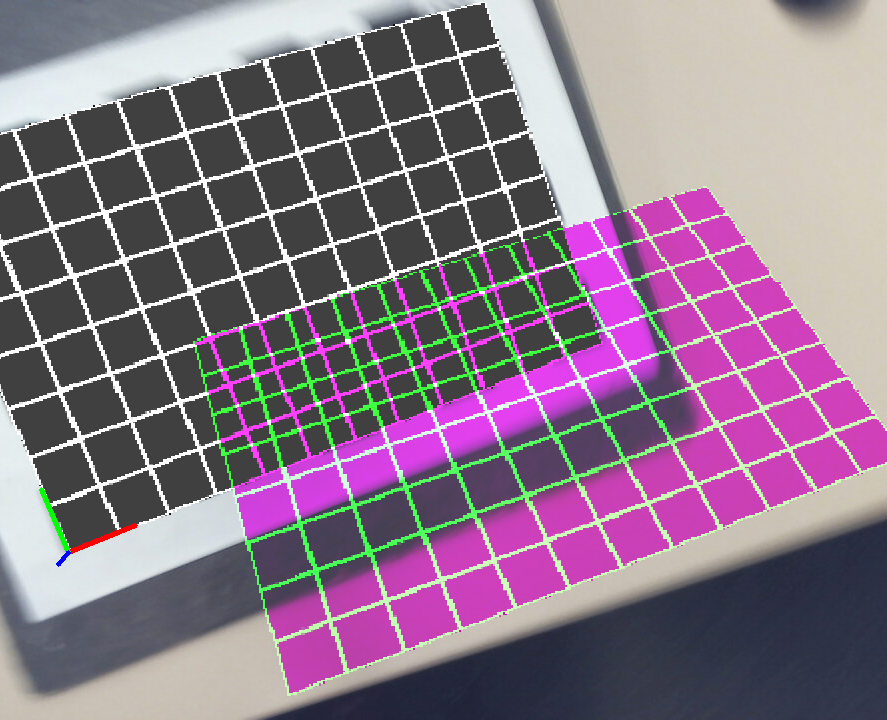}
}
\caption{The two overlay patterns we evaluated in our second user survey. Note that we now do overpaint the real board in the video stream.}
\label{fig:posevis}
\end{figure}

Complementing the first survey, the second survey determined whether one can take advantage of the geometric property that the pattern can be flipped horizontally and vertically.
To this end we chosen two different visualization of the calibration pattern as follows:

\begin{enumerate}
\item The asymmetric chessboard as in used for the preceding survey.
\item A quadrille paper visualization, which is fully symmetric yet still contains the necessary perspective cues.
\end{enumerate}

To keep the connection between the target pose overlay and the physical calibration board when using the new visualization, we overpaint the actual calibration target in the video stream - similarly to \citep{richardson2013iros}.
We also apply the over-painting to the first option (see figure \ref{fig:posevis}) to exclude the effect of tracking imprecision from the survey.

There were 7 Participants in this survey which had to reach 10 target poses. As with the preceding survey the order of the options was randomized so 3 participants started with option 1 and 3 participants started with option 2.

We only used the "virtual mirror" setup based on the results from the first survey.

\section{Results}

In the following the results of our user surveys are shown. First we discuss the quantitative timings of each experiment. Then we also present some qualitative observations made during the trials.

\subsection{Quantitative results}

Table \ref{tbl:view} shows the quantitative results of the first survey, giving the per-user times as well as the overall average. 

\begin{table}
\centering
\begin{tabular}{|c|c|c|}
\hline
& t (first-person view)  & t (virtual mirror)  \\ 
\hline
\hline
User 1 & 1:39 min & 1:44 min\\ 
\hline
User 2 & 3:17 min & 1:08 min\\
\hline
User 3 & 2:46 min & 1:55 min\\ 
\hline
User 4 & 7:22 min & 1:36 min\\ 
\hline
User 5 & 2:22 min & 1:25 min\\ 
\hline
\hline 
Mean & 3:29 min & 1:33 min \\ 
\hline 
\end{tabular} 
\caption{time the users required to reach 9 given target poses.}
\label{tbl:view}
\end{table}

The average calibration time of 1:33 min to complete the calibration show a strong advantage of the virtual mirror scenario over the first-person view approach with an average time of 3:29 min. Looking at the individual results we see that only User 1 is slightly faster using the first-person view, while all other Users were considerably faster using the virtual mirror approach. User 4 even struggles to complete the calibration using in the first-person view.
Therefore we conclude that the virtual mirror approach is preferable.

\begin{table}
\centering
\begin{tabular}{|c|c|c|}
\hline
& t (chessboard)  & t (quadrille paper)  \\ 
\hline
\hline
User 1 & 2:14 min & 1:00 min \\
\hline
User 2 & 2:07 min & 1:20 min \\
\hline
User 3 & 3:06 min & 2:11 min \\
\hline
User 4 & 3:43 min & 3:20 min \\
\hline
User 5 & 1:21 min & 1:44 min \\
\hline
User 6 & 1:52 min & 2:00 min \\
\hline
\hline 
Mean & 2:24 min & 1:56 min \\ 
\hline 
\end{tabular} 
\caption{time the users required to reach 10 given target poses with different visualizations}
\label{tbl:pattern}
\end{table}

Table \ref{tbl:pattern} shows the results of the second survey, again giving the average as well as the per user times. There are only 6 results given as one participant failed to match the first intrinsic pose within ~3 min with any method. Therefore we aborted the trial and no results are given.

The average time of 1:56 min to complete the calibration using the quadrille paper visualization shows a slight advantage over the chessboard visualization with 2:24 min. However looking at the individual results there are 2 participants being faster using the chessboard visualization. Furthermore there is strong variation between the individual users. Therefore no clear conclusion can be given. 

\subsection{Qualitative results}

Additionally to the times presented above we made the following qualitative observations:

\begin{itemize}
\item It took the participants much longer to match the intrinsic pose then the distortion pose.
\item With the "quadrille paper" pattern, some users did not rotate the pattern to match the distortion calibration pose, but rather moved it out of view.
\item The users reached a target pose faster if it was from the same category as the previous one; e.g. if a distortion pose followed a distortion pose. 
Conversely the needed to re-orient if e.g. a distortion pose followed a intrinsic pose.
\item When asked about the experience users preferred the "quadrille paper" visualization - even if their calibration time was higher in this mode.
\end{itemize}

Here the time it took the participants to match the intrinsic pose was the determining factor in overall calibration time.

\section{Conclusion}
We have presented an evaluation of different user guidance methods for camera calibration. This allows us to give a recommendation that the "virtual mirror" setup is preferable for camera calibration. However the results of our second survey only hint that using the simplified "quadrille paper" overlay is of advantage. While the user feedback was generally positive and we measured a slight advantage in the average calibration time, there was a strong variation between the individual participants.
Therefore a larger scale survey is necessary to give a definitive answer here.

However our qualitative observations indicate that larger gains are to be expected from adapting the pose sequence then from modifying the pattern visualization. 
Particularly the arbitrary switching between the pose categories requires physical and mental switching on the user side.
Additionally we observed that matching 3 arbitrary rotations of the pattern to the target pose took considerably longer then to match the position.

Therefore the pose sequence should adapted to further improvements on user guidance. Currently the poses optimize the algorithmic constrains while neglecting the user. One option would be to find for a better compromise between these two objectives.
Alternatively one could introduce "guidance only" poses that are placed between the current pattern position and the target pose. Those would not be used for calibration, but rather to give the user more hints on how to reach the target pose. Trivially one could insert the neutral pose between two calibration poses s.t. only 3 DOF change between each two displayed targets. 

\bibliographystyle{plain}
\bibliography{bibliography}

\end{document}